\documentclass[aps,twocolumn, prl, showpacs]{revtex4}
\usepackage{amsmath}
\usepackage{graphicx}

\begin{document}

\title{Power Spectra of the Total Occupancy
in the Totally Asymmetric Simple Exclusion Process}
\author{D.A. Adams, R.K.P Zia, and B. Schmittmann}
\date{May 15, 2007}


\begin{abstract}
As a solvable and broadly applicable model system, the 
totally asymmetric exclusion process enjoys iconic status in the 
theory of non-equilibrium phase transitions.
Here, we focus on the time dependence of the total number of particles 
on a 1-dimensional open lattice, and its power spectrum. 
Using both Monte Carlo simulations and analytic methods, we explore 
its behavior in different characteristic regimes. In the 
maximal current phase and on the coexistence line (between high/low 
density phases), the power spectrum displays algebraic decay, with exponents 
$-1.62$ and $-2.00$, respectively. Deep within the high/low density phases, 
we find pronounced \emph{oscillations}, which damp into power laws. This behavior 
can be understood in terms of driven biased diffusion with conserved noise in the bulk.
\end{abstract}

\affiliation{Department of Physics, Virginia Tech, Blacksburg, VA 24061-0435 USA}
\maketitle

\emph{Introduction.} The collective behavior of many constituents,
interacting under \emph{non-equilibrium }conditions, is far from well
understood. Yet, such systems are ubiquitous in nature, from molecular
biology at the nanoscale to infractructure networks at the global level. In
physics, attacks on such highly complex systems often begin with seemingly
small steps, defining simple models that are both tractable and believed to
capture the essentials of the original problem. On this stage, the totally
asymmetric simple exclusion process (TASEP) \cite{TASEP} plays a key role,
much like the Ising model in equilibrium statistical mechanics. With open
boundaries, particles stochastically enter a one-lane lattice from one end,
hop to the next site if it is empty, and leave at the opposite end. This
simple transport model provides the first crucial steps towards the modeling
of realistic processes, such as protein synthesis, surface growth, and
vehicular traffic \cite{TASEP-appl}. Further, it generates much theoretical
interest: Despite its simplicity, its stationary state displays several
phases and interesting algebraic correlations, much of which is known analytically \cite{TASEP-sol}.

In this study, we focus on another simple quantity in TASEP: the total
number of particles in the system at time $t$, $N(t)$. In the steady state,
its time average should be the ensemble average, which can be easily
computed from the known stationary distribution. However, as a fluctuating
quantity, its power spectrum, $I(\omega )$, contains time-correlation
information which is not as easily accessible. Although such correlations
have been investigated recently \cite{PPOF}, we find interesting behavior
undetected previously: \emph{oscillatory behavior} in $I(\omega )$ for the
high and low density phases. While the previous study concerns correlations
within the bulk of an \emph{infinite} system, the new feature here is that
our $I\left( \omega \right) $ carries information on the \emph{entirety} of
a \emph{finite} system. Since many physical systems are far from the
thermodynamic limit (e.g., mRNAs containing around $10^{3}$ codons or
fewer), such finite-size effects can be physically significant. In the
remainder of this letter, we briefly summarize the details of the model, our
simulation methods, report our findings, and provide theoretical
explanations for the phenomena.

\emph{The model and simulation results.} A standard TASEP consists of
particles hopping on a lattice of length $L$ with site label $i$. If the
first site is empty, a particle enters the system with rate $\alpha $. A
particle on the last site leaves the system with rate $\beta .$ In the bulk,
a particle always hops onto the next site (with unit rate) if the site is
empty; otherwise, it remains stationary. A configuration at any particular
time is described by a set of occupation numbers $\left\{ n_{i}\right\} $ ($%
n=0,1$ for a vacant vs. occupied site). In all our simulations, we start
with an empty lattice and use random sequential updates, i.e., in each Monte
Carlo step (MCS), we make $N+1$ random attempts to move a particle. Since
the system is stochastic, a complete description requires $P\left( \left\{
n_{i}\right\} ,t\right) $, the probability to find the system in
configuration $\left\{ n_{i}\right\} $ at time $t$. The evolution of $P$ is
governed by a simple master equation. Though linear and typically easy to
write, this equation cannot be solved in general. Yet, in the long time
limit, $P$ settles into a $t$-independent distribution: $P^{\ast }$, the
exact form of which is known \cite{TASEP-sol}.

\begin{figure}[t]
\begin{center}
\resizebox{85mm}{!}{\includegraphics{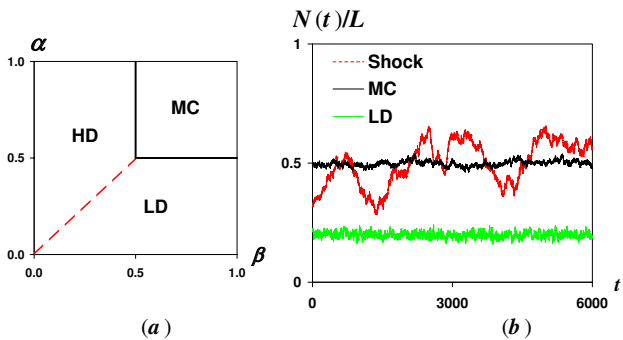}}
\end{center}
\par
\vspace{-0.7cm} 
\caption{(a) Phase diagram of TASEP. On the dashed line, high
and low density domains coexist, with a microscopic interface, known as the
``shock.'' (b) Typical time series (up to 600K MCS shown) of $N\left(
t\right) $, for $L=1000$, in the three distinct regimes: 
coexistence, MC and LD (color online). }
\vspace{-0.5cm}
\label{fig:PhaseDiagram}
\end{figure}

The macroscopic properties of these stationary states can be categorized in
terms of three distinct phases, conveniently displayed as a phase diagram
(Fig.~\ref{fig:PhaseDiagram}a) in the $\alpha $-$\beta $ plane. Thanks to
the exact solution, $P^{\ast }$, many properties can be computed
analytically. The three regions are associated with the high density (HD),
low density (LD), and maximum current (MC) phases. They are distinguished by
their average local densities, $\rho_{i}\equiv \left\langle
n_{i}/L\right\rangle $, where $\left\langle \bullet \right\rangle $ is an
average over the stationary $P^{\ast }\left( \left\{ n_{i}\right\} \right) $%
. In the thermodynamic limit and deep in the bulk, $\rho_{i}$ is $\rho
_{+}=1-\beta $, $\rho _{-}=\alpha $, and ${1/2}$ for, respectively, the HD,
LD, and MC phases. All aspects of the LD and HD phases can be directly
related through particle-hole symmetry, so that we will concentrate only on
the LD phase in this letter. 
For the LD phase, the deviations from $\rho _{-}$ are
confined to a microscopic boundary layer near the exit. In the MC phase, the
profile decays algebraically into the bulk, while the system displays
behavior comparable to critical phenomena in equilibrium statistical
mechanics. The transitions at the boundaries of the MC phase are continuous.
By contrast, a discontinuity occurs across the HD/LD line.
On this line ($\alpha =\beta <1/2$), 
typical configurations display the
coexistence of regions with $\rho _{+}$ 
and $\rho _{-}$, joined by a
microscopically sharp interface, known as a ``shock''. 
The shock is delocalized, performing a random 
walk over the entire lattice, so that the average 
$\rho_{i}$ is linear over the interval $\left[ \alpha ,1-\alpha %
\right] $.

Our main interest here is $N(t)\equiv \sum_{i}n_{i}(t)$ and its associated
power spectrum: $I(\omega )$. Though its time average (in the steady state)
is just $\sum_{i}\left\langle n_{i}\right\rangle \equiv \bar{\rho}L$, $%
I(\omega )$ cannot be accessed from $P^{\ast }$. Specifically, we take a
measurement of $N$ every 100 MCS (after discarding the first 100K MCS to
allow the system to reach steady state) and label these by $t=1,2,...,T$.
Typical $N(t)$'s in the three regimes are shown in Fig.~\ref%
{fig:PhaseDiagram}b. For all but a few of our runs, $T=10^{4}$ ($10^{6}$
MCS), so that we can define a Fourier transform: $\tilde{N}\left( \omega
\right) \equiv \sum_{t=1}^{T}N(t)e^{i\omega t}$, where $\omega =2\pi m/T$
with $m=0,1,...$ . To obtain the average power spectrum, we carry out
typically 100 such runs and construct 
\begin{equation}
I(\omega )=\left\langle \left| \tilde{N}\left( \omega \right) \right|
^{2}\right\rangle .  \label{eq:PSFormula}
\end{equation}%
The important control parameters for this system are $L$, $\alpha $, and $%
\beta $. We investigated a range of $L$'s from $250$ to $32000$. Here, we
will show the results for $L=1000$, as well as some for the largest size.
Also, we will present mainly data for three representative points in the
phase diagram. Labeled by $\left( \alpha ,\beta \right) $, they are $\left(
0.7,0.7\right) $, $\left( 0.3,0.7\right) $, and $\left( 0.3,0.3\right) $,
corresponding to, respectively, the MC, LD phases and the coexistence line.
Systematic studies of the rest of the $\alpha $-$\beta $ square will be
reported elsewhere \cite{promises}.


\begin{figure}[tb]
\begin{center}
\resizebox{70mm}{!}{\includegraphics[angle=0]{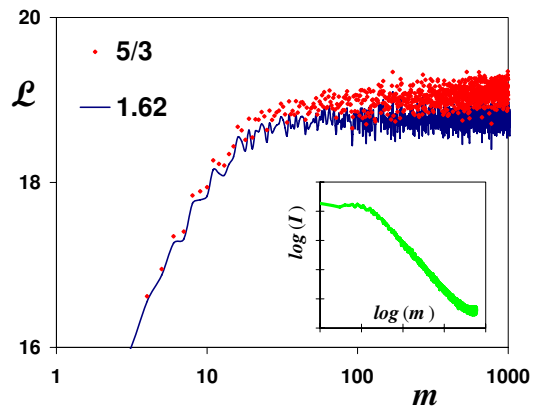}}
\end{center}
\par
\vspace{-0.7cm}
\caption{Power spectrum in the MC phase, 
$\protect\alpha = \protect\beta =0.7$. with $L=1000$. 
The inset shows a log-log plot of 
$I\left( \protect\omega \right) $. The main figure shows 
$\mathcal{L}\equiv \ln \left( \protect\omega ^pI\right)$ 
vs. $ m \equiv \protect\omega T/2 \protect\pi $ (color online).}
\label{fig:MC_Spectra}
\vspace{-0.5cm}
\end{figure}

\begin{figure}[tb]
\begin{center}
\resizebox{70mm}{!}{\ \includegraphics[angle=0]{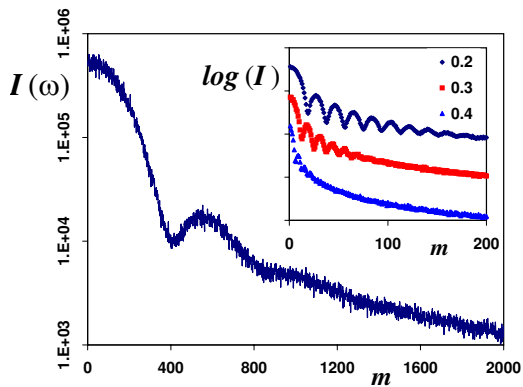}}
\end{center}
\par
\vspace{-0.5cm} 
\caption{Power spectrum $I \left( \protect\omega \right)$  
vs. $m\equiv \protect\omega T/2\protect\pi$ in the LD phase 
($\protect\alpha =1-\protect\beta =0.3$) with $L=1000$. 
The inset shows a similar plot for a $L=32000$ lattice,
with $\protect\alpha =1-\protect\beta =0.2,\,0.3,\,$ and $0.4$. 
The units for $log \left( I \right)$ are arbitrary, and the 
curves are displaced for clarity (color online).}
\vspace{-0.5cm}
\label{fig:LD_Spectra}
\end{figure}

In the inset of Fig.~\ref{fig:MC_Spectra}, we show the power spectra for the
MC phase. Away from high and low $\omega $ values, $I\left( \omega \right) $
appears to obey a power law: $I\propto \omega ^{-p}$. The saturation at
small $\omega $ is due to finite $L$ and the crossover scales with the
expected $L^{-3/2}$ (see below). At large $\omega $, we are
presumably probing an ultraviolet cutoff, due to a discrete lattice and MCS.
To expose the power $p$ more effectively, we display $\omega ^{p}I$ in a
log-log plot (Fig.~\ref{fig:MC_Spectra}). The best fit seems to be $p=1.62$
(line, blue online), though the data are still consistent with $5/3$ (dots,
red online), a value favored by the theoretical considerations below.  Whether the difference of 
$\sim 3\%$ is truly significant remains to be explored \cite{promises}.
We have also simulated other points in the MC phase and find all data 
to be statistically indistinguishable from the set here.
The results for the coexistence line, at $\alpha =\beta =0.3$, are very
similar, except that they confirm the expected random walk behavior, 
with a power of $2$. The most remarkable phenomena are found in the 
HD/LD phases. As illustrated in Fig.~\ref{fig:LD_Spectra}, 
$I\left( \omega \right) $ displays large, 
\emph{damped oscillations}. The examples in
the inset are especially striking: An $L=32K$ lattice is used here, with
three $\alpha $'s. In the next section, we briefly present the
theory which accounts for these properties.

\emph{Theoretical approaches.} To understand the behavior of the power
spectra above, we rely on different (though related) approaches for the
three regimes. The simplest case is the coexistence line, where the shock
performs a random walk. As a result, $N\left( t\right) $ is also just a
random walk 
(Fig.~\ref{fig:PhaseDiagram}b) 
in the interval $\left[ \rho _{-}L,\rho _{+}L\right] $ with
reflecting boundary conditions \cite{DW}. The power spectrum of this process
is well understood, displaying the $\omega ^{-2}$ behavior. Of course, for
small $\omega $ (and long times), $I\left( \omega \right) $ saturates to a
finite value, controlled by the allowed interveral for $N$. Later \cite%
{promises}, we will show that the expected crossover exhibits data collapse
with a scaling variable $\omega L^{2}$. The next regime, considered the most
challenging theoretically, is the MC phase. Here, the associated
non-equilibrium steady state is ``critical'', displaying (power law)
correlations and anomalous exponents. At the coarse-grained level, our
system is described by a continuous density, $\rho \left( x,t\right) $, and
may be studied as a stochastic field theory. Indeed, the TASEP is a driven
diffusive system \cite{DDS} in one dimension. Deviations of $\rho $ about
its average here, $\phi \left( x,t\right) \equiv \rho \left( x,t\right) -1/2$%
, satisfy the noisy Burgers equation \cite{Burgers}: $\partial _{t}\phi =%
\frac{1}{2}\partial _{x}^{2}\phi +\partial _{x}\phi ^{2}+\partial _{x}\eta $%
, where $\eta $ is a Gaussian noise. Within the bulk of an infinite system,
the dynamic two-point function $\left\langle \phi \left( x,t\right) \phi
\left( x^{\prime },t^{\prime }\right) \right\rangle $ is translationally
invariant and assumes a scaling form \cite{PPOF,DDS,FNS}: 
\begin{equation}
C\left( \xi ,\tau \right) =\xi ^{-1}f\left( \xi /\tau ^{2/3}\right) 
\label{scaling}
\end{equation}%
where $\xi =x^{\prime }-x$ and $\tau =t^{\prime }-t$. If we naively consider 
$\int d\xi C\left( \xi ,\tau \right) $ as a candidate for our power
spectrum, we would find that Eqn. (\ref{scaling}) leads to $\int dsf\left(
s\right) $, \emph{independent of }$\tau $! This surprising conlusion is
consistent with simply setting $k=0$ in the results of \cite{JS}. On closer
examination, however, we should seek the Fourier transform of $%
\int_{0}^{L}dxdx^{\prime }\left\langle \phi \left( x,0\right) \phi \left(
x^{\prime },t\right) \right\rangle $ in a \emph{finite }system. If we
approximate this two-point function by $C\left( \xi ,\tau \right) $, we are
led to $\int_{0}^{L}d\xi \left[ 2\left( L-\xi \right) C\left( \xi ,t\right) %
\right] $, which is of the form $t^{2/3}g\left( L/t^{2/3}\right) $. Thus,
its Fourier transform would take the scaling form $I\left( \omega \right)
=\omega ^{-5/3}G\left( L\omega ^{2/3}\right) $. Assuming that $G$ approaches
a positive constant for $\omega \gg L^{-3/2}$, this theory predicts $I\left(
\omega \right) \propto \omega ^{-5/3}$. As discussed above, our data are
consistent with this power, although $1.62$ appears to be a better fit.
Though a 3\% disagreement seems minor, we found similar discrepancies in
other observables such as 
$\left\langle N\left( t\right) N\left( t^{\prime }\right) \right\rangle $. These issues, along with finite-size corrections to
scaling, discrete space and time, and possible systematic errors, will be
explored elsewhere \cite{promises}.

Typically, the behavior deep in the HD and LD phases seems least
interesting, 
with ordinary Gaussian fluctuations.
Yet, 
$I\left( \omega \right) $ for this regime exhibits the most structure. We
now turn to a brief account of a simple, linear theory for the fluctuations
that predicts these interesting oscillations, focusing on the LD phase only.

Following standard routes for a continuum description \cite{DDS,PPOF} of a
driven density $\rho (x,t)$, we study small fluctuations, $\phi $, around $%
\bar{\rho}$ using the stochastic equation of motion: $\partial _{t}\phi
(x,t)=D\partial _{x}^{2}\phi -v\partial _{x}\phi +\partial _{x}\eta $. Here, 
$D$ is the (effective) diffusion constant, $v$ is the bias, and $\eta $ is a
Gaussian correlated (local) noise with zero mean. If we were to start from
the microscopic $\left\{ n_{i}\right\} $ and take a naive continuum limit,
we would arrive at $D=1/2$ and $v=1-2\bar{\rho}$ (in units of lattice
spacing and MCS). But we leave these as $O\left( 1\right) $ parameters for
now. The solution in Fourier space is easy:\ $\tilde{\phi}\left( k,\omega
\right) =ik\tilde{\eta}\left( k,\omega \right) /\left[ Dk^{2}+ivk-i\omega %
\right] $, where $\phi (x,t)=\int_{k,\omega }\tilde{\phi}\left( k,\omega
\right) e^{i(kx-\omega t)} $ and $\int_{k}\equiv \int dk/2\pi $. 
With $N\left( t\right) =\int_{0}^{L}dx\rho \left( x,t\right) =%
\bar{\rho}L+\int_{0}^{L}\phi \left( x,t\right) $, we have $\tilde{N}\left(
\omega \right) =\int_{k}\left( e^{ikL}-1\right) \tilde{\eta}/\left[
Dk^{2}+ivk-i\omega \right] $ for $\omega >0$. Writing $\left\langle \tilde{%
\eta}\left( k,\omega \right) \tilde{\eta}^{\ast }\left( k^{\prime },\omega
^{\prime }\right) \right\rangle =A\delta \left( k-k^{\prime }\right) \delta
\left( \omega -\omega ^{\prime }\right) $ and replacing $\delta \left(
\omega -\omega \right) $ by $T$, we find the power spectrum to be $%
\frac{AT}{2\pi }\int_{k}\left| \left( e^{ikL}-1\right) /\left[
Dk^{2}+ivk-i\omega \right] \right| ^{2}$. Evaluting the integral, we arrive
at

\begin{equation}
I\left( \omega \right) =\frac{8ADT}{v^3}\mathop{\rm Re}\left[ \frac{%
e^{ik_{+}L}-1}{R\left( 1-R\right) ^2}+ \frac{e^{ik_{-}^{*}L}-1}{R^{*}\left(
1+R^{*}\right) ^2}\right]  \label{eq:N2}
\end{equation}
where $R\equiv \sqrt{1-4iD\omega /v^2}$ and $k_{\pm }=iv\left( -1\pm
R\right) /2D$.

\begin{figure}[tb]
\begin{center}
\resizebox{70mm}{!}{\includegraphics[angle=0]{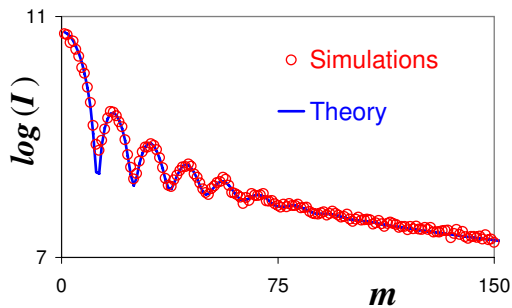}}
\end{center}
\par
\vspace{-0.5cm} 
\caption{Power spectrum $\log \left( I\right)$ 
vs. $m$ for $\protect\alpha =1-\protect\beta =0.3$ with
$L=32000$ (color online).} 
\vspace{-0.5cm}
\label{fig:fit32K}
\end{figure}

To shed light on this complex expression, we consider three frequency
regimes: (a) $\omega \rightarrow 0$, e.g., $m=1,2,3...$, (b) intermediate $%
\omega $'s ($\omega \ll v^{2}/D$, i.e., $1\ll m\ll T$), and (c) $\omega
\rightarrow \infty $. In regime (a), $I$ approaches a constant, which can be
used to fit $A$. A further simplication is that the first term in $%
\mathop{\rm Re}\left[ ...\right] $ is $O\left( \left( vL/D\right)
^{2}\right) $, while the second term is $O\left( 1\right) $. Thus, the
latter can be largely ignored, if we take, e.g., $L=32K$, as in the inset of
Fig.~\ref{fig:LD_Spectra}. In regime (b), we keep $O\left( \omega
^{2}\right) $ corrections in $R$ and $k_{\pm }$, i.e., 
\begin{equation*}
1-\frac{2iD\omega }{v^{2}}+\frac{2D^{2}\omega ^{2}}{v^{4}}\quad \text{and}%
\quad \frac{\omega }{v}+\frac{iD\omega ^{2}}{v^{3}}
\end{equation*}%
respectively. The result shows damped oscillations manifestly: 
\begin{equation}
I\left( \omega \right) \cong \frac{2AvT}{D\omega ^{2}}\left[ 1-e^{-\frac{%
D\omega ^{2}L}{v^{3}}}\cos \left( \omega L/v\right) \right] .
\label{eq:SmallOmegaFinal}
\end{equation}%
The minima are near integer multiples of $2\pi v/L$ in $\omega $ (i.e., $vT/L
$ in $m$). Remarkably, the minima in Fig.~\ref{fig:LD_Spectra} are very
close to these values if we naively substitute $\left(
0.4,10^{6},10^{3}\right) $ for $\left( v,T,L\right) $. Encouraged by this
approach, we attempt to fit the $L=32K$ data with the full Eqn. 
(\ref{eq:N2}). The result for $\alpha=0.3$, shown in Fig.~\ref{fig:fit32K}, 
is surpringly good, indicating that this simple theory has captured the 
essentials of our system.

The physical origins of the oscillations are noteworthy. They can be traced
to the time it takes a fluctuation to traverse the entire lattice. Indeed,
if there is no exclusion and each particle travels ballistically at velocity 
$u$, then $N\left( t\right) $ will consist of a random series of unit
increments (due to $\alpha $), correlated with unit drops at times $L/u$
later (assuming $\beta =1$). The associated $I$ will be proportional to $%
z^{-2}\sin ^{2}z$, with $z\equiv \omega L/2u$, with zeros at multiples of $%
2\pi u/L$. The effect of diffusion, embodied in $D$, is just to smear out
the minima and fill in the zeros. Returning to the fit above, we were
surprised that it requires $D\simeq 20$ and $A\simeq 1/500$, values that
seem remarkably far from the naive $O\left( 1\right) $ levels. Work is in
progress to understand why these parameters are so seriously
``renormalized.''\cite{promises}

Beyond the oscillations, there is another subtle feature in this
intermediate $\omega $ regime. As Eqn. (\ref{eq:SmallOmegaFinal}) shows, the
oscillations are completely damped when $\omega ^{2}$ exceeds $v^{3}/DL$,
and $I$ settles into $\omega ^{-2}$. Though not shown, this power is
confirmed in the $L=32K$ case. However, on further scrutiny, we see that $%
R\rightarrow \sqrt{\omega }$ in the truly asymptotic regime (c): $\omega \gg
v^{2}/4D$. Then, Eqn. (\ref{eq:N2}) predicts a $\omega ^{-3/2}$ decay \cite%
{3/2 power}! If this regime is reached \emph{before} the oscillations are
fully damped, then the $\omega ^{-2}$ behavior of the ``intermediate
regime'' will lie hidden. This is indeed true for the $L=1K$ case,
where the oscillations die out at much larger $m$'s and only the $3/2$ power
is observed.

\emph{Outlook.} In addition to a more detailed and systematic Monte Carlo study (e.g.,
larger systems, longer runs, finite-size scaling, etc.), there are a number
of issues worthy of further pursuit. The continuum theory represents a rough
first step and should be refined to a version with discrete space/time. Most
intimately connected to this work is a better understanding of the origin of
the large ``renormalization effects'' on the diffusion coefficient, $D$.
Beyond interests specific to the simple TASEP, we are motivated by its
applicability to protein synthesis, where the lattice (particles) model the
mRNA (ribosomes) \cite{TASEP-appl}. To capture more of the process \emph{in
vivo}, we should include the effects of (i) particles 
that occupy $\ell > 1$ sites, 
(ii) inhomogeneous hopping rates, 
(iii) finite reservoir of particles which can enter the lattice, and 
(iv) competition with other TASEPs (mRNAs) for a finite pool of particles. 
Some of these issues have been considered \cite{TASEP-appl}, 
though
none focused on the total occupancy, a quantity surely of interest in the
context of biology. Needless to say, many other generalizations
- all well motivated by physical systems - 
come to mind, e.g., many particle species, multiple
lanes, and high dimensions.  
In the
past, power spectra have served us well, providing valuable perspectives
into stochastic systems in general \cite{PSpGenRef} and driven system in
particular \cite{PSpDDS}. We hope this work will spark further interest to
study power spectra in novel non-equilibrium systems far
beyond physics, such as biology, social networks, and finance.


\begin{thebibliography}{99}
\bibitem{TASEP} F. Spitzer, Adv. Math. \textbf{5}, 246 (1970). For a recent
review, see G. Sch\"{u}tz, \emph{Exactly solvable models in many-body systems%
}, Vol. 19 of \emph{Phase Transition and Critical Phenomena}, eds C. Domb
and J.L. Lebowitz (Academic, London, 2001).

\bibitem{TASEP-appl} C. MacDonald, J. Gibbs, and A. Pipkin, Biopolymers, 
\textbf{6}, 1 (1968); C. MacDonald and J. Gibbs, Biopolymers, \textbf{7},
707 (1969); D.E. Wolf, and L.-H. Tang, Phys. Rev. Lett. \textbf{65}, 1591
(1990); V. Popkov, L. Santen, A. Schadschneider, and G.M. Sch\"{u}tz, J.
Phys. A: Math. Gen. \textbf{34}, L45 (2001); L.B. Shaw, R.K.P. Zia, and K.H.
Lee, Phys. Rev. E \textbf{68}, 021910 (2003); T. Chou and G. Lakatos, Phys.
Rev. Lett. \textbf{93}, 198101 (2004).

\bibitem{TASEP-sol} See, e.g., B. Derrida, Phys. Rep. \textbf{301}, 65
(1998) and the review in \cite{TASEP}.

\bibitem{PPOF} P. Pierobon, A. Parmeggiani, F. von Oppen, and E. Frey, Phys.
Rev. E 72, 036123 (2005). Note that the power spectrum considered in this
reference (Eqn. 26) is associated with the \emph{autocorrelation} of the
local density fluctuations. It should be distinguished from our correlation
of the\emph{\ total} \emph{occupation} in the system.

\bibitem{promises} D.A. Adams, B. Schmittmann, and R.K.P. Zia, to be
published.

\bibitem{DW} B. Derrida, E. Domany, and D. Mukamel, J. Stat. Phys. \textbf{69%
}, 667 (1992). L. Santen + C. Appert J. Stat. Phys. \textbf{106}, 187 (2002).

\bibitem{DDS} H.K. Janssen and B. Schmittmann, Z. Phys. B\textbf{63}, 517
(1986); For a general review, see B. Schmittmann and R.K.P. Zia \emph{%
Statistical Mechanics of Driven Diffusive Systems}, Vol. 17 of \emph{Phase
Transitions and Critical Phenomena}, eds C. Domb and J.L. Lebowitz
(Academic, New York, 1995).

\bibitem{Burgers} J.M. Burgers, \emph{The Nonlinear Diffusion Equation }%
(Riedel, Boston, 1974).

\bibitem{FNS} D. Forster, D.R. Nelson, and M.J. Stephen, Phys. Rev. A\textbf{%
16}, 732 (1977).

\bibitem{JS} Specifically, see Eqns. (21b) and (24a) in \cite{DDS} and also %
\cite{FNS} and H. van Beijeren, R. Kutner, and H. Spohn, Phys. Rev. Lett. 
\textbf{54}, 2026 (1985).

\bibitem{3/2 power} We emphasize that, despite the similarity, this $I \sim
\omega ^{-3/2}$ is \emph{not} the same as in \cite{PPOF}, which focused on
the power spectrum of the \emph{local} density fluctuations.

\bibitem{PSpGenRef} See, e.g., H. L. Pecseli, \emph{Fluctuations in Physical
Systems }(Cambridge, UK, 2000).

\bibitem{PSpDDS} See, e.g., J.V. Andersen, H.K. Jensen, and O.G. Mouritsen, 
Phys. Rev. B\textbf{44}, 439 (1991); K-t. Leung, 
Phys. Rev. B\textbf{44}, 5340 (1991); K.B. Lauritsen and H.C. Fogedby, 
J. Stat. Phys. \textbf{72}, 189 (1993).
\end{thebibliography}
\end{document}